\documentclass[twocolumn,pre,showpacs,preprintnumbers,amsmath,amssymb]{revtex4}

\usepackage{graphicx}
\usepackage{dcolumn}
\usepackage{bm}
\usepackage{color}
\usepackage{amsmath}
\definecolor{b}{rgb}{0,0,1.0}
\definecolor{r}{rgb}{1,0,0}
\definecolor{g}{rgb}{0,1,0}

\begin{document}

\preprint{APS/pre-print}

\title{Polymeric filament thinning and breakup in microchannels} 

\author{P.E. Arratia$^{1}$}
\author{J.P. Gollub$^{1,2}$}
\author{D.J. Durian$^{1}$}%
\affiliation{%
$^{1}$Department of Physics and Astronomy, University of
Pennsylvania, Philadelphia, PA 19104\\$^{2}$Department of Physics,
Haverford College, Haverford, PA 19041
}%

\date{\today}

\begin{abstract}
The effects of elasticity on filament thinning and breakup are investigated in microchannel cross flow. When a viscous solution is stretched by an external immiscible fluid, a low 100~ppm polymer concentration strongly affects the breakup process, compared to the Newtonian case. Qualitatively, polymeric filaments show much slower evolution, and their morphology features multiple connected drops. Measurements of filament thickness show two main temporal regimes: flow- and capillary-driven. At early times both polymeric and Newtonian fluids are flow-driven, and filament thinning is exponential. At later times, Newtonian filament thinning crosses over to a capillary-driven regime, in which the decay is algebraic. By contrast, the polymeric fluid first crosses over to a second type of flow-driven behavior, in which viscoelastic stresses inside the filament become important and the decay is again exponential. Finally, the polymeric filament becomes capillary-driven at late times with algebraic decay. We show that the exponential flow thinning behavior allows a novel measurement of the extensional viscosities of both Newtonian and polymeric fluids.

\end{abstract}

\pacs{47.50.-d, 47.55.df, 83.50.Jf}
\maketitle

\section{\label{sec:level1}Introduction}

The progressive breakup of an initially stable fluid thread into small drops or bubbles is a rich phenomenon of great interest~\cite{Eggers1997}. For example, flow focussing in microfluidic devices can continuously produce drops or bubbles whose sizes are controlled by the relative flow rate of the two immiscible fluids \cite{Anna.Stone2003, Tabeling2003, Gordillo.Weitz2004, Garstecki.Stone2004, Link.Stone2004, Garstecki.Whitesides2005}. While most such work concerns Newtonian fluids, many fluids of interest for lab-on-a-chip applications are likely to exhibit complex micro-structure and non-Newtonian behavior, such as viscoelasticity. Furthermore, viscoelastic effects, which can be quantified by the Elasticity number \emph{El}=$\lambda$$\eta$/($\rho$$L^{2}$), scale inversely with the square of the device length scale (\emph{L}), and are likely to be accentuated in microfluidic devices. Here, $\lambda$ is the fluid relaxation time, $\eta$ is viscosity, and $\rho$ is density. For polymeric drop breakup in macroscopic flow, elasticity can give rise to breakup behavior that is markedly different from that of Newtonian fluids~\cite{Goldin.Shinar1969, Wagner.Eggers2005, Clasen.McKinley2006, Tirtaatmadja.Cooper-White2006}. For example, a viscoelastic filament driven by gravity in a quiescent bath~\cite{Entov.Hinch1997} undergoes an initial linear viscous decrease in the filament diameter, followed by a slower thinning process in which capillary forces are balanced by the fluid elastic stresses.

Recently, a numerical investigation in a flow-focusing device~\cite{Zhou.Feng2006} showed qualitative differences with respect to Newtonian fluids such as prolonged thinning of the fluid filament and delay of drop pinch-off. No measurements of thinning rates or breakup times were presented. An experimental investigation in a `T' shaped geometry using a low viscosity, elastic fluid~\cite{Hunsy.Cooper-White2006} also found prolonged thinning of the fluid filament. The authors observed a linear decrease in filament diameter followed by a `self-thinning' exponential regime, which was argued to have a rate inversely proportional to the fluid relaxation time ($\lambda$). However, $\lambda$ was found to vary over an order of magnitude with shear rate, though it should remain constant. While both investigations found similar qualitative trends, no quantitative connection has yet been made to the extensional flow within the filament during thinning and breakup.

In this paper, we compare the filament thinning and breakup of Newtonian and viscoelastic fluids of equal shear viscosity in a microchannel cross-slot geometry.  Here, the outer Newtonian fluid stretches the inner Newtonian or polymeric fluid into a thin filament until it eventually breaks up into drops. This geometry allows for very fine control of the flows over a broad range of shear rates. Measurements of filament thickness show two temporal
regimes: (i) a flow-driven regime in which the filament thins exponentially and (ii) a capillary-driven regime in which the filament thins algebraically. Our analysis leads to a novel method of measuring the steady extensional viscosities of both Newtonian and polymeric fluids.

\begin{figure}
\includegraphics[scale=0.45]{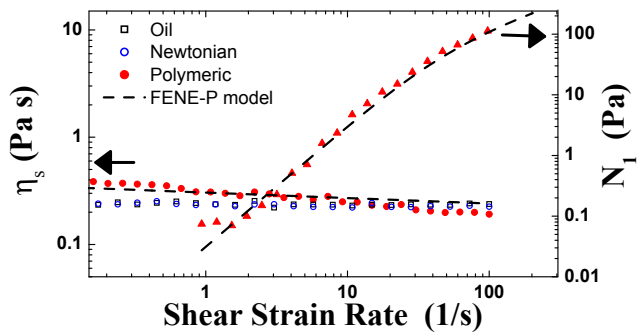}
\caption{\label{rheo} (Color Online) Fluid rheological
characterization. (Left y-axis) Shear viscosity vs shear strain rate for
all fluids; oil=mineral oil; Newtonian=water/glycerin mixture;
polymeric=PAA in water/glycerin mixture. The shear viscosity is
nearly constant even for the polymeric solution
$\eta_{s}\approx0.24$~Pa s. (Right y-axis) First normal stress
difference for the polymeric solution vs shear strain rate. Dashed curves
represent fits using the FENE-P model with parameters
$\lambda=0.45$~s and $b=4500$.}
\end{figure}

\section{\label{sec:level1}Methods}

The experimental configuration is a cross-slot microchannel, $W=50~\mu$m wide and $L=30~\mu$m deep, molded in poly(dimethylsiloxane) (PDMS, Dow Sylgard 184) using standard soft-lithography methods~\cite{Quake2000,Sia.Whitesides2003}. Channels are sealed with a glass cover slip after exposure to an oxygen plasma. In order to keep the microchannel wetting properties uniform, the glass cover slip is coated with a thin layer of PDMS prior to the exposure. The assembled channels are then baked for 12~hrs at 100 $^{\circ}$C in order to obtain hydrophobic walls wetted by the continuous outer liquid phase.

The outer continuous phase is mineral oil containing 0.1\% by weight of surfactant (SPAN 80, Fluka). Both Newtonian and polymeric fluids are used for the inner (or ``dispersed") phase. The Newtonian fluid is a 90\%-glycerin aqueous solution. The polymeric fluid is made by adding 100 ppm of high molecular weight polyacrylamide (PAA, $M_W = 18 \times 10^6$, 15\% polydispersity), which has a flexible backbone, to a Newtonian 85\%-glycerin aqueous solution with a measured shear viscosity of $\eta_{s,solv}=0.2$~Pa s; the water/glycerin mixture is used as a solvent for the polymer. It is dilute, below the overlap concentration of approximately 350 ppm. The interfacial tension between the continuous and dispersed phases is $\sigma=10$~mN/m. The fluids are characterized with a stress-controlled rheometer at 25 $^{\circ}$C. As shown in Fig.~\ref{rheo}, the shear viscosities of the oil and Newtonian fluids are nearly identical and independent of shear strain rate: $\eta_{s} \approx 0.24$~Pa s. Also as shown, the viscoelastic polymeric fluid exhibits nearly constant shear viscosity (power law index=0.97) and a first normal stress difference $N_{1}$, which increases with shear strain rate.

We fit the polymeric fluid shear rheology data to the widely-used finite extensibility nonlinear elastic model with Peterlin's closure (FENE-P)~\cite{Peterlin1966,Bird1987,McKinley.Shinar2002}. In this model the fluid total stress tensor $\tau$ is assumed to be the sum of a contribution from the solvent and another resulting from the presence of polymer molecules such that $ \tau=\tau_{solv}+\tau_{poly}$. The solution shear viscosity $\eta_s$ is then the sum of the solvent and polymeric parts $ \eta_{s}=\eta_{s,solv}+\eta_{s,poly}$. The FENE-P model is well adapted for dilute (and semidilute) high molecular weight polymeric solutions, and has been used previously to analyze filament thinning of polymeric fluids in macroscopic experiments~\cite{Wagner.Eggers2005}. A fluid described by the FENE-P model possesses the same dynamical properties as a fluid described by the much simpler Oldroyd-b model~\cite{Bird1987}, which assumes that polymers can be modeled as Hookean springs. The main difference is that the Oldroyd-b model allows for infinite extension of polymer molecules, while the FENE-P model uses a spring-force law in which the polymer molecules can be stretched only by a finite amount in the flow field~\cite{Peterlin1966,Bird1987}.

\begin{figure}
\includegraphics[scale=2.0]{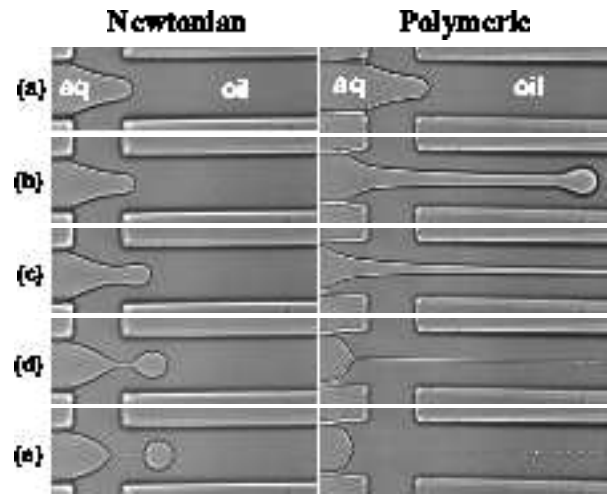}
\caption{\label{photos}  Evolution of the thinning process for Newtonian (left column) and polymeric fluids (right column), for a flow rate ratio \emph{q}=\emph{Q$_{oil}$}/\emph{Q$_{aq}$}=60, where \emph{Q$_{oil}$}/\emph{Q$_{aq}$} corresponds to the ratio of oil and aqueous phase flow rates. Oil is the continuous (outer) phase while the aqueous phase is either Newtonian or polymeric.  (a)Initial regime; (b) t/t$_{b}$ = 0.15, where t$_{b}$ is breakup time; (c) t/t$_{b}$ =0.45; (d) t/t$_{b}$ = 0.95; (e) after breakup. Values of t$_{b}$ for the Newtonian and polymeric cases are 11.5 ms and 245 ms, respectively. Note the appearance of satellite droplets in the Newtonian case and multiple beads attached to the filament in the polymeric case (d,e). The channel width and depth are 50 $\mu$m and 30 $\mu$m, respectively.}
\end{figure}

A simultaneous fit (Fig.~\ref{rheo}) of the polymeric fluid $\eta_{s}$ and $N_{1}$ data to the FENE-P model provides the fluid relaxation time $\lambda$ and a dimensionless finite extensibility parameter $b$, which are the only two adjustable parameters~\cite{Bird1987}. The best fit results in $\lambda=0.45$~s and $b=4500$. Further details on the equations and methods used to fit the FENE-P model to the shear rheology can be found elsewhere~\cite{Lindner.Bonn2003}.

The dispersed and continuous phases are injected into the central and side arms of the cross-channel, respectively, using syringe pumps (Harvard Instruments). Experiments are performed for flow rate ratios, $q=Q_{oil}/Q_{aq}$, ranging from 10 to 200. In all cases, the aqueous flow rate is kept constant at $Q_{aq}=0.01$~l/min.  This is low enough that the behavior is quasi-static, such that the periodicity -but not the morphology- depends on $Q_{aq}$. For this range of parameters, the Reynolds number is less than 0.01; therefore viscous forces are much larger than inertial forces. Similarly the capillary number ranges from 0.02 to 0.8; therefore, viscous forces are also larger than surface forces.  Under these conditions an aqueous filament is formed and stretched by the flow of the surrounding oil.  The thinning and breakup of the filament are imaged using an inverted microscope and a fast video camera, with frame rates between 1 and 10~kHz.

\section{\label{sec:level1}Observations}
\subsection{Qualitative}

Sample frames from video data are shown in Fig.~\ref{photos}, for both Newtonian and polymeric fluids, at a flow rate ratio of $q=60$. The Newtonian case, shown in the \emph{left-column}, displays typical filament thinning and drop formation. The aqueous phase is drawn into the cross-slot channel (a), and begins to elongate and collapse (b-d), forming a primary drop connected to a very thin filament; later (e) the filament thins at a faster rate and breaks into a large primary drop and small satellite droplets.

The polymeric case, shown in the \emph{right-column} of Fig.~\ref{photos}, displays very different behavior.  Initially (a), we observe a morphology that is similar to that of the Newtonian fluid, i.e. viscoelasticity is negligible at first. As the thinning progresses, the polymeric fluid develops a longer neck with a drop attached to it (b). This filament elongates while thinning at a slower rate than in the Newtonian case (c). Near the breakup event, the polymeric fluid shows multiple beads (`beads-on-a-string') attached to the filament (d) \cite{Goldin.Shinar1969,Clasen.McKinley2006,Chang.Kalaidin1999}. After breakup, there are many satellite drops (e).

\begin{figure}
\includegraphics[scale=1.1]{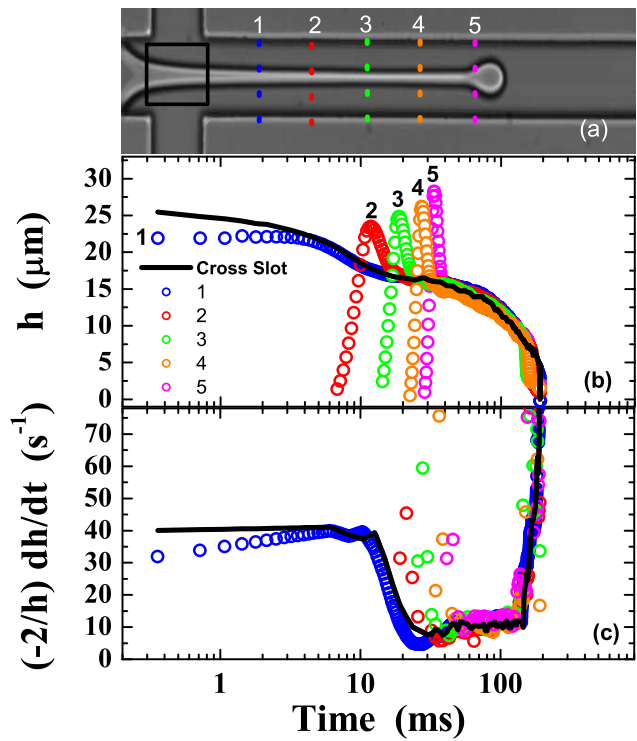}
\caption{\label{cwdt} (Color Online) Position independence of the measurement. (a) Filament thickness $h(t)$ measured at different locations in the microchannel. Measurements are performed in the cross-slot region (solid line rectangle) and at (dashed lines) 1, 2, 3, 4, and 5 channel widths downstream from that point. (b) The filament thickness for the polymeric fluid at $q$=60 measured at different positions. Data is color coded according to (a). (c) The computed extensional strain rate $\dot{\varepsilon}$ for the same cases shown in (b). The data shows that the measurements of $h(t)$ are nearly independent of axial position, after an initial transient.}
\end{figure}

\subsection{Quantitative}

The filament thinning process is quantified by the decrease in diameter $h(t)$ as a function of time.  To accomplish this, we fit a third-order polynomial equation to the interface contour, which is restricted to the cross-slot region. The field of view corresponding to the cross-slot region, in which $h(t)$ measurements are performed, is delimited by the solid line rectangle shown in Fig.~\ref{cwdt}(a).  We assume that the interface is symmetric across the centerline and only half of the contour is fitted with the polynomial. We then locate the absolute value of the minimum in the polynomial first derivative. The filament diameter is measured at the point where the absolute value of the minimum in the first derivative is located. There are instances, however, where the minimum in absolute slope may be located at edge of the cross-slot region. Hence, we must check the dependence of $h(t)$ on measurement location, i.e. axial position $z$.

We test the dependence of $h(t)$ on axial position $z$ by measuring $h(t)$ in the cross-slot region and also 1, 2, 3, 4, and 5 channel widths downstream from the edge of the cross-slot region (Fig.~\ref{cwdt}a). Results are presented in Fig.~\ref{cwdt}(b); the values of $h(t)$ measured at different locations in the channel are nearly the same except for an initial transient. It follows that the values of the extensional strain rate $\dot{\varepsilon}$ (Fig.~\ref{cwdt}c) measured at different locations are also very similar. Here, we assume that $\dot{\varepsilon} = -(2/h){\rm d}h/{\rm d}t$. We will check the validity of this assumption next.

The extensional strain rate can be assumed to be $\dot{\varepsilon} = -(2/h){\rm d}h/{\rm d}t$ only if the filament thickness $h$ is homogeneous in the axial coordinate $z$~\cite{Amarou.Bonn2001,Oliveira.McKinley2005}. However, there is some variation with $z$ and an extra term in the extensional strain rate that is proportional to $({\rm d} h/{\rm d} z)$ may arise. In order to check whether this extra term can be neglected (or not), we consider an argument based on dimensional analysis: to convert $(v/h)({\rm d} h/{\rm d} z)$ to a strain rate requires an inverse timescale, which must be given by a speed over a length. The only speeds in the system are $v$ and ${\rm d} h/{\rm d} t$. Here, $v$ is the average fluid velocity inside the filament, which is much larger than ${\rm d} h/{\rm d} t$.  The only lengths in the system are $h$ and the channel width, $W$; the former is smaller.  Therefore the biggest possible extra term in the extensional strain rate $\dot{\varepsilon}$ would be a constant times $(v/h)({\rm d} h/{\rm d} z)$.

\begin{figure}
\includegraphics[scale=1.05]{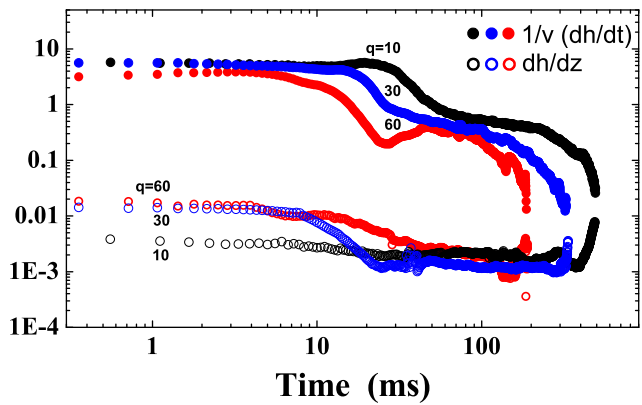}
\caption{\label{dhdz} (Color Online) Filament axial spatial gradient ${\rm d} h/{\rm d} z$ and normalized extensional strain rate $(1/v) {\rm d} h/{\rm d} t$ as a function of time for a polymeric fluid filament.  Here, $v$ is the average velocity inside the fluid filament. Measurements are performed for different flow rate ratios $q$=10, 30, and 60. The data shows that ${\rm d} h/{\rm d} z$$<<$$(1/v) {\rm d} h/{\rm d} t$ so filament thickness spatial gradients may be neglected when computing the extensional strain rate $\dot{\varepsilon}$.}
\end{figure}

Following the argument above, we compare the space and time derivatives (Fig.~\ref{dhdz}). We express them non-dimensionally as ${\rm d} h/{\rm d} z$ and $(1/v){\rm d} h/{\rm d} t$, where the prefactor ($1/v$) makes the time derivative dimensionless. We find that the space derivative of the filament thickness is at least an order of magnitude smaller than the dimensionless time derivative. Hence, the extensional strain rate can be safely assumed to be $\dot{\varepsilon} = -(2/h){\rm d}h/{\rm d}t$.

To summarize, the results in Fig.~\ref{cwdt} and Fig.~\ref{dhdz} show that one can, to a good approximation, study the thinning process by treating the filament as if it is nearly uniform spatially, with a thickness that depends only on time.

\begin{figure}
\includegraphics[scale=1.2]{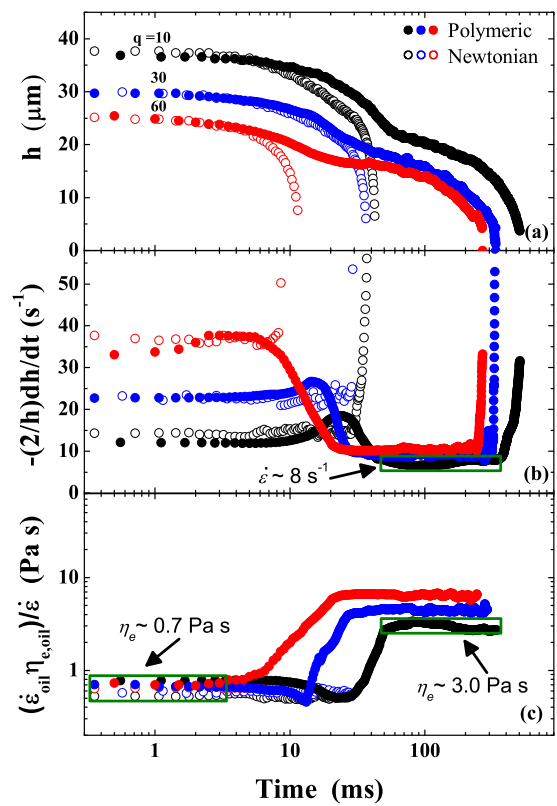}
\caption{\label{hvst} (Color Online) Time-dependent filament thinning (a) Filament thickness \emph{h}(t) for both Newtonian and polymeric fluids for \emph{q}=10, 30, and 60. (b) Filament extensional strain rate $\dot{\varepsilon}$=-(2/\emph{h})d\emph{h}/d\emph{t} for the same fluids. Both viscous and elastic regimes are characterized by constant $\dot{\varepsilon}$. The value of $\dot{\varepsilon}$ is equal to 8$~s^{-1}$ for a polymeric fluid at $q$=10. (c) The quantity $(\dot{\varepsilon}_{oil}\eta_{e,oil})/\dot{\varepsilon}$ is the filament extensional viscosity $\eta_{e}$ where the flow is extensional, i.e., $\dot{\varepsilon}=constant$. Here, $\dot{\varepsilon}_{oil}$ and $\eta_{e,oil}$ are the oil extensional strain rate and extensional viscosity, respectively. Initially, for all fluids, the values of $\eta_{e}$ are similar since all fluids have nearly the same $\eta_{s}$. Later, the asymptotic value of $\eta_{e}$ increases with larger values of $q$.}
\end{figure}

\section{\label{sec:level1}Results}
\subsection{Flow-Driven Regime}

In Fig.~\ref{hvst}(a), we present sample results of measurements of filament thickness $h(t)$, performed in the cross-slot region, as a function of time. We show data for both Newtonian and polymeric fluids for three flow rate ratios, $q=10$, 30, and 60. At short times, the Newtonian and polymeric fluids exhibit identical initial thinning, which is indicative of their common $\eta_{s}$. But at longer times, the two diverge with the polymeric filament lasting at least an order of magnitude longer before breakup. We also note shorter breakup times as \emph{q} is increased. This trend is also found in other flow-focusing experiments~\cite{Thorsen.Quake2001,Anna.Stone2003} and in a numerical investigation~\cite{Zhang.Stone1997} using Newtonian fluids.

The filament extensional strain rate $\dot{\varepsilon} = -(2/h){\rm d}h/{\rm d}t$ is shown as a function of time for the same flow rate ratios $q$, in Fig.~\ref{hvst}(b). For the Newtonian fluid, $\dot\varepsilon$ is initially independent of time; therefore, in this regime, $h(t)$ decreases exponentially with time. For the polymeric fluid, $\dot\varepsilon$ is initially equal to the same constant as for the Newtonian fluid.  But it soon departs and, after a transient interval, settles down to smaller constant value, which indicates a second regime of slower exponential thinning.  For all fluids at the very latest times, close to breakup, the final decrease of $h(t)$ to zero gives an apparent divergence of $\dot\varepsilon$. We show in Section D that the data just before breakup are consistent with a linear decrease in filament diameter, $h(t)\propto(t-t_b)$ where $t_b$ is the breakup time.

To model the exponential decrease of filament diameter, we assume that (1) filament thinning is driven mainly by the outer fluid extensional flow in the cross-slot region and (2) the shear flow that develops is relatively far downstream from the cross-slot region and should have no implications on the local stress balance. These are reasonable assumptions since shear stresses tangential to the filament do not contribute to the thinning (or squeezing) of the filament; filament thinning is driven by viscous stresses normal to the filament.

Starting from an assumption of stress balance inside and outside the interface, and applying the definition of extensional viscosity \cite{McKinley.Shinar2002}, we obtain the condition $\eta_{e}\dot{\varepsilon}=\eta_{e,oil}\dot{\varepsilon}_{oil}$, which relates the strain rates and extensional viscosities of the inner and outer phases. Here, the left and right sides are the extensional viscosity multiplied by the extensional strain rate for the aqueous filament and continuous oil phases, respectively. As discussed above, the strain rate in the filament is $\dot{\varepsilon} = -(2/h){\rm d}h/{\rm d}t$. The strain rate for oil in the cross-slot region is $\dot{\varepsilon}_{oil} \approx Q_{oil}/(W^2L)$, as verified by particle-tracking methods \cite{Arratia.2006}. Lastly, since the oil
is Newtonian, its extensional viscosity is $\eta_{e,oil}=3\eta_{s,oil} $, where $\eta_{s,oil}$ is the oil shear rate viscosity \cite{McKinley.Shinar2002,Trouton1906}. Therefore, also assuming that $\eta_e$ is independent of time, the filament diameter thins exponentially according to

\begin{equation}
   h(t)=h_o \exp[-(3/2)(\eta_{s,oil}/\eta_e)\dot{\varepsilon}_{oil}t].
\label{hexp}
\end{equation}
where $h_o$ is an integration constant.This equation is valid for the two flow-driven exponential regimes shown in Fig.~\ref{hvst}. In such {\it flow-driven} regimes, Eq.~(\ref{hexp}) may be used to deduce $\eta_e$ from $h(t)$ data.

We note that the quantity $\dot{\varepsilon}_{oil}$ is measured in the cross-slot region, where the flow is extensional and where pinching from the `mother drop' occurs. To this end, we have checked that $\dot{\varepsilon}_{oil}$ remains constant during the filament thinning and breakup event; the average velocity of the oil in the cross-slot region is constant.

The transition between the two exponential thinning regimes can be elucidated by plotting the quantity $\varphi=(\dot{\varepsilon}_{oil}\eta_{e,oil})/\dot{\varepsilon}$, which has units of viscosity, as a function of time (Fig.~\ref{hvst}c). We find that $\varphi$ is nearly constant in regions where $\dot\varepsilon$ is constant. In such regions, the quantity $\varphi$ is the same as the filament extensional viscosity $\eta_e$.

The values of $\eta_e$ are computed for each steady extensional strain-rate $\dot{\varepsilon}$, which is proportional to $q$, as shown in Fig.~\ref{hvst}(c). We find that i) the initial value of $\eta_e$ is independent of $q$ and ii) the asymptotic value of $\eta_e$ increases as the flow rate ratio $q$ is increased. In macroscopic experiments~\cite{McKinley.Shinar2002}, the values of steady values of $\eta_e$ are also known as the asymptotic extensional viscosity, which corresponds to a state where polymer chains are fully stretched. In this investigation, however, asymptotic extensional viscosity means the degree of extension of polymer chains in the fluid filament for a given value of $\dot{\varepsilon}$.

\begin{figure}
\includegraphics[scale=0.63]{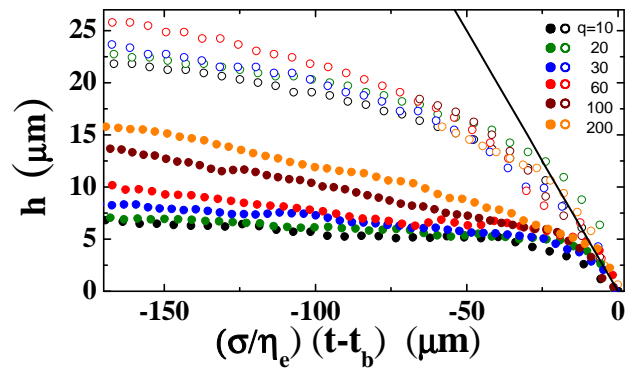}
\caption{\label{capillary}(Color Online) Filament thickness versus time in the capillary driven regime, where $t_{b}$ is the breakup time. At very late times, the filament thins roughly linearly in time with a speed proportional to $\sigma/\eta$$_{e}$ for both Newtonian (open circles) and polymeric (filled circles) fluids. The flow ratio (\emph{q}) is color-coded in the legend for both cases. The solid line represents slope=-1/2.}
\end{figure}

\subsection{Capillary-Driven Regime}
The linear decrease of the filament thickness near the final breakup can also be modeled by stress balance, now by incorporating surface tension effects. Specifically, the Rayleigh-Plateau instability eventually sets in so that capillary forces cause beading and ultimately breakup. Equating radial stress with the Laplace pressure gives $\eta_e\dot{\varepsilon} = \sigma / h$~\cite{Cohen.Nagel1999,Zhang.Lister1999,Garstecki.Whitesides2005}. Therefore, the filament diameter thins linearly with time according to

\begin{equation}
    h(t) = -(1/2)(\sigma/\eta_e)(t-t_b),
\label{hlin}
\end{equation}
where $t_b$ is the breakup time. In such {\it capillary-driven} regimes, Eq.~(\ref{hlin}) may be used to deduce $\eta_e$ from $h(t)$ data. Equation 2 shows that, near the singularity, $h(t)$ varies linearly with $(t-t_b)$ with slope $\sigma/2\eta_{e}$, which has been observed numerically~\cite{Lister.Stone1998} in the Stokes regime, except that, in the numerical work, shear rather than extensional viscosity is used in the denominator.

To demonstrate the consistency of our extensional viscosity results in the flow- and capillary-driven regimes, we plot data for $h(t)$ vs
$(\sigma/\eta_e)(t-t_b)$ in Fig.~\ref{capillary}. There, the value of $\eta_e$ is taken from analysis of the flow-driven regime using Eq.~(\ref{hexp}). To within apparently random deviations, the $h(t)$ data vanish linearly with $(\sigma/\eta_e)(t-t_b)$ with slope $-1/2$, in accord with Eq.~(\ref{hlin}).  Note however that the dynamic range is limited, since the imaging resolution is about $2~\mu$m. Therefore, the capillary-driven regime is consistent with the flow-driven regime, but the latter gives more accurate values of extensional viscosity $\eta_e$.

\section{Discussion}

The extensional properties of polymeric fluids are important for applications such as turbulent drag reduction and splash suppression
\cite{McKinley.Shinar2002,Berg.Bonn2000}. However, measurement of $\eta_e$ has remained a difficult task~\cite{Anna.James2001}. We now show that high-quality data on the values of steady extensional viscosity for both polymeric and Newtonian fluids can be obtained using our method.

\begin{figure}
\includegraphics[scale=1.1]{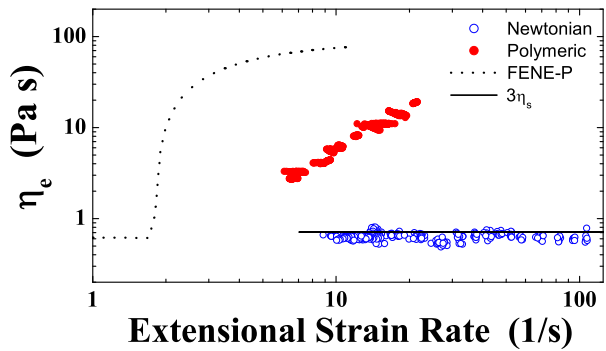}
\caption{\label{extvis} (Color Online) Extensional viscosities of both Newtonian and polymeric fluids, derived from the filament
thinning measurements and Eq.~(\ref{hexp}), as a function of the extensional strain rate $\dot{\varepsilon}$=-2/\emph{h}(d\emph{h}/d\emph{t}). The polymeric fluid extensional viscosity shows strain hardening and increases with a power law exponent of approximately 1.0.  The theoretical Trouton ratio of a Newtonian fluid is 3.0 (solid line). The FENE-P model prediction is also shown, but is far from the measurements.}
\end{figure}

Final results for $\eta_e$ based on Eq.~(\ref{hexp}) are plotted in Fig.~\ref{extvis} vs extensional strain rate. Here each point represents a different fixed flow-rate ratio, $q$. For the Newtonian fluid, $\eta_e$ is independent of extensional strain rate and nearly equals $3\eta_s$ as expected \cite{McKinley.Shinar2002,Trouton1906}.  This agreement serves as a second check, complementary to Fig.~\ref{capillary}. For the polymeric fluid at early times, in the first flow-driven regime, the behavior is the same as for the Newtonian fluid (not shown). At later times, in the second flow-driven regime, the extensional strain rate of the filament is lower and $\eta_e$ is higher. This `strain hardening' behavior is due to the stretching of the polymer molecules in the extensional flow of the thinning filament, and it has been observed in other macroscopic experiments \cite{Amarou.Bonn2001,Anna.McKinley2001}.

It is important to point out that the values presented in Fig.~\ref{extvis} are for steady extensional viscosity and not transient extensional viscosity, which is usually reported in macroscopic experiments~\cite{Amarou.Bonn2001,Anna.McKinley2001}. Here, values of $\eta_e$ are computed for each steady extensional strain-rate $\dot{\varepsilon}$, which is proportional to $q$, as shown in Fig.~\ref{hvst}(c) and Fig.~\ref{extvis}. In macroscopic experiments, the values of asymptotic $\eta_e$ are measured when polymer chains are fully stretched, while here the asymptotic $\eta_e$ means the degree of extension of polymer chains in the fluid filament for a given value of $\dot{\varepsilon}$.

In Fig.~\ref{rheo}, the FENE-P model properly describes both the $\eta_s$ and $N_{1}$ versus shear rate with two adjustable parameters, which are $\lambda$=0.45 s and \emph{b}=4500. An expression for $\eta_e$ can be obtained from the FENE-P model for a range of extensional strain rates~\cite{Lindner.Bonn2003,Bird1987} using the values of  $\lambda$, $b$, and $\eta_{s,solv}$. The FENE-P prediction for $\eta_e$ is plotted in Fig.~\ref{extvis}. It exhibits strain-hardening behavior, which saturates at high strain rates by accounting for the finite extensibility of the polymer molecules. However, by comparison with our data, the predicted strain hardening sets in too soon and too abruptly. A possible source of error in the model may be polymer dispersivity ($\sim$15\% in M$_{W}$), which can smear out the sharp rise in $\eta_{e}$~\cite{Wagner.Eggers2005}. It cannot, however, account for such early transition to strain hardening behavior since $\lambda$ $\sim$ M$_{W}$$^{3/2}$.

Other sources of error may be the inherent limitations of the FENE-P model such as the averaging of the force values connecting the beads in the dumb-bell model originally proposed by Peterlin \cite{Peterlin1966}. This averaging is known to lead to unexpectedly large polymeric stresses compare to the non-averaged FENE model \cite{vanHeel1998}. Another limitation is that while real polymeric fluids have a spectrum of $\lambda$, the FENE-P model, as used here, is described by a mean $\lambda$ obtained in a shear flow, which is known to be low for use in extensional flows.  Therefore, we should expect some type of failure of predictions of $\eta_{e}$ based on the single mode FENE-P model. This disagreement does not imply a weakness in the measurement.

\section{Conclusion}

In conclusion, small amounts of flexible polymer can dramatically affect filament thinning and breakup in microchannel extensional
flow. In contrast to macroscopic observations, we find both a \emph{flow-driven} regime in which the filament thins followed by a
\emph{capillary-driven} regime responsible for filament breakup. For a Newtonian fluid, the filament thins exponentially with time until onset of capillary surface tension-induced breakup. For the polymeric fluid with the same shear viscosity (nearly independent of shear strain rate), there is an intermediate regime in which the filament thins exponentially at a much slower rate.  Furthermore in the capillary regime a series of small droplets is generated along the filament. These differences may be attributed solely to extensional viscosity and its increase with extensional strain rate, since this is the only rheological difference between the Newtonian and polymeric fluids. For thinner filaments and faster thinning, the polymer molecules stretch and cause an increase in extensional viscosity without significant change in shear viscosity.

Measurements of the exponential rate of thinning can thus be used to determine the steady extensional viscosity, an elusive quantity to measure. For
the Newtonian case, $\eta_e\approx~3\eta_s$; for the polymeric case, the values of $\eta_e$ increase with extensional strain rate, but much more slowly than predicted by the FENE-P model. This suggests the need for a better understanding of both the molecule-scale behavior of polymers in extensional flows as well as its connection to macroscopic rheology. Filament thinning in microchannels, and its variations with polymer molecular weight, may be a promising approach.

\section{Acknowledgments}

We thank Daniel Bonn, Gareth McKinley, and Howard Stone for fruitful discussions. Seth Fraden and Katie Humphry provided help with microfabrication
methods. Kerstin Nordstrom and Ben Polak provided assistance with experiments. This work was supported the National Science Foundation through grant MRSEC/DMR05-20020.


\bibliography{polymer_drop_refs_long}

\end{document}